\documentclass[a4paper]{article}

\usepackage{INTERSPEECH2022}

\usepackage{color}
\usepackage{enumitem}
\usepackage{epstopdf}
\usepackage{multirow}
\usepackage{url}
\usepackage[skip=2pt]{caption}
\usepackage[table]{xcolor}

\definecolor{myb}{RGB}{96, 125, 163}
\definecolor{myy}{RGB}{212, 189, 47}
\colorlet{myo}{orange}
\colorlet{myr}{red}

\title{Assessing ASR Model Quality on Disordered Speech using BERTScore}
\name{Jimmy Tobin\textsuperscript{1}, Qisheng Li\textsuperscript{2}, Subhashini Venugopalan\textsuperscript{1}, Katie Seaver\textsuperscript{1,3}, Richard Cave\textsuperscript{1,4}, Katrin Tomanek\textsuperscript{1}}
\address{
  \textsuperscript{1}Google Research, USA; \\
  \textsuperscript{2}University of Washington, USA; \\
  \textsuperscript{3}Leonard Florence Center for Living, USA; \\
  \textsuperscript{4}MND Association, UK}
\email{\{jtobin,katrintomanek\}@google.com}

\begin{document}

\maketitle
\begin{abstract}
Word Error Rate (WER) is the primary metric used to assess automatic speech recognition (ASR) model quality. It has been shown that ASR models tend to have much higher WER on speakers with speech impairments than typical English speakers. It is hard to determine if models can be be useful at such high error rates. This study investigates the use of BERTScore, an evaluation metric for text generation, to provide a more informative measure of ASR model quality and usefulness. Both BERTScore and WER were compared to prediction errors manually annotated by Speech Language Pathologists for error type and assessment. BERTScore was found to be more correlated with human assessment of error type and assessment. BERTScore was specifically more robust to orthographic changes (contraction and normalization errors) where meaning was preserved. Furthermore, BERTScore was a better fit of error assessment than WER, as measured using an ordinal logistic regression and the Akaike's Information Criterion (AIC). Overall, our findings suggest that BERTScore can complement WER when assessing ASR model performance from a practical perspective, especially for accessibility applications where models are useful even at lower accuracy than for typical speech.
\end{abstract}

\section{Introduction}
\label{sec:intro}
Automatic Speech Recognition (ASR) model quality is most often measured by Word Error Rate (WER), the aggregate score of word deletions, substitutions and insertions. WER is the \textit{de facto} metric for benchmarking models' improvements and regressions. Most state-of-the-art (SOTA) ASR models trained and evaluated on typical English speech report single digit WERs on standard test corpora \cite{synnaeve_2021}. However, models trained only on typical speech perform worse on disordered speech, reporting a median WER of 31.5 on short phrases~\cite{green2021}. In these situations where WER is high, it is difficult to assess how useful a model will be for the end user. In particular, WER doesn't capture semantic closeness or the assessment of the error. For instance, misrecognizing the suffix `s' in a plural and missing the negation of a word can have the same error rate but missing the negation can fail to capture the user's intent (see examples in Table ~\ref{tab:error_examples}). Thus, looking at word error rate alone may not convey the utility of the model in some practical settings where errors are tolerated. An example being video call captioning for a person with a speech impairment, which can ensure that the conversation partner will understand the speaker better.

Green et al. show that, for individuals with disordered speech, personalizing ASR models with the individual's speech can lead to on average 75\% relative WER improvements over unadapted models trained for typical speech~\cite{green2021}. The personalized models, although having WERs in the range of 15 or 20 (which would be considered high for typical speech), provided utility to the users in certain domains such as home automation, spoken transactions (asking for something), and conversations \cite{tobin2022}. Their work demonstrates that transcriptions do not need to be perfect to communicate a speaker’s intent or meaning. Errors such as normalization (one hundred percent vs. 100\%) or contraction expansion (I'm vs I am) can be considered a combination of deletions, insertions and substitutions, but to a human or a SOTA natural language understanding (NLU) system, the meaning is conveyed the same. 

Alternatives to WER have been investigated, aiming to address the limitations when applied to conversations~\cite{Wang2003,morris04_interspeech}. There are several metrics in natural language understanding applications that aim to measure similarity of generated text with a human truth e.g. BLEU~\cite{Papineni_2002} and METEOR~\cite{banerjee-lavie-2005-meteor} are popular in machine translation, ROUGE~\cite{lin-2004-rouge} is used the context of summarization, SPICE~\cite{Anderson_spice} is used in image-captioning, and more recently BLEURT~\cite{sellam-etal-2020-bleurt} and BERTScore~\cite{zhang2019bertscore} which are based on pre-trained language models have been proposed for text generation.
Even for speech applications, recent work~\cite{kim2021, sharma2022end} proposed metrics utilizing pre-trained language models. In Kim et al.~\cite{kim2021}, the proposed alternate metric is based on semantic distance between the predicted and ground truth transcripts. Our work is motivated similarly, though we utilize BERTScore\footnote{BERTScore is available open source at https://github.com/Tiiiger/bert\_score.}, an existing semantic distance based metric popular in language generation and adapt it to our task. %

In this work, we specifically investigate BERTScore and WER for measuring the utility of personalized ASR models. We worked with Speech Language Pathologists (SLPs) to perform a detailed analysis of the errors, including error type and assessment assessments, based on the model predictions from personalized ASR models. We then compared  both BERTScore and WER in terms of how well they aligned with the SLP assessments. Our findings showed that while both metrics are well correlated with SLP judgements, BERTScore was a better fit to the assessments overall. Further, BERTScore was a better indicator of utility in the presence of simple contraction and normalization errors than WER.

\section{Data and Approach}
\label{sec:methods}
We first describe the model and data used in our analysis and then describe the metric and comparison approach.

\subsection{Personalized ASR Models and utility}
The ASR models used in our study are based on work from Green et. al.~\cite{green2021} which developed personalized models for 432 speakers with impairments who contributed to Project Euphonia~\cite{macdonald2021}. This corpus consists of over 1 million samples (over 1300 hours) of more than
1000 anonymized speakers with different types and severity
levels of speech impairments.
Following the same approach in \cite{green2021}, we fine-tuned models on a subset of 15 individuals from the Euphonia corpus~\cite{macdonald2021} by tuning most of the initial encoder layers while keeping the decoder frozen. This included speakers having varying degrees of speech impairment described in Table~\ref{tab:severity_counts}. Speaker etiologies include Amyotrophic Lateral Sclerosis (ALS), Cerebral Palsy, Down Syndrome, Multiple Sclerosis, and other long tail etiologies that have less representation in the overall Euphonia dataset, including Brain Injury, Vocal Chord Paralysis and Cleft Palate. This approach showed huge improvements in WER, from average WER of 66.3 $\pm$ 10.0 before adaptation to a WERs in the range of 15-20 after personalization. These WERs though, are still much higher than the original model's WER of 6-8 on typical speech~\cite{Prabhavalkar_min_wer_2018}. 

\begin{table}[h]
\centering
\begin{tabular}{|l|l|l|}
    \hline
    \multirow{2}{*}{\textbf{Severity}} & \multirow{2}{*}{\textbf{\# Speakers}} & \textbf{Avg. Adapted WER } \\
    & & (rel. improvement) \\
    
    \hline
    Mild & 2 & 16.5 (62\%)\footnotemark \\
    Moderate & 7 & 14.3 (76\%) \\
    Severe & 6 & 21.6 (72\%)  \\
    \hline
  \end{tabular}
  \caption{Distribution of speakers, severity of speech impairment, and average WER after adaptation with relative improvement.}
  \label{tab:severity_counts}
\end{table}
\footnotetext{Mild speech severity models having a higher average WER than moderate speech severity is an artifact of the small sample size. Green et al. (2021) showed that \textit{Mild} WER was significantly lower than \textit{Moderate} in the Euphonia dataset. }

To validate the utility of the models beyond the WER improvements, particularly for overall user experience, we solicited feedback from 15 speakers who used personalized ASR models to dictate their speech through a mobile app for several months. 
The feedback suggested that in several scenarios, the models were indeed useful. Specifically, users were able to (1) perform home automation through voice controlled assistants, (2) use it for face to face and video conversations with friends and family, (3) long-form dictation and email, and (4) even for short transactional interactions with individuals unfamiliar with the impaired speaker's voice. 
These user success stories suggested that despite high WERs, users experienced benefits across several domains and situations.

\subsection{Error Analysis}
To understand and measure the utility of the models and track improvements and regressions, %
we sought more detailed analysis of the errors from SLPs on the per-speaker test set produced by each speaker's personalized model.
SLPs assessed a total of 3473 transcription errors and determined (1) error type and (2) assigned an assessment rating of the error severity. 
They categorized the errors into 8 types including deletion, contraction, normalization, homophone, spelling, proper noun, repetition, and other word errors. These are described in Table~\ref{tab:error_types}. 
The error severity assessment was rated on a scale of 0, 1 or 2 depending on whether or not an error preserved meaning. This is described in Table~\ref{tab:error_assessment scale}. %
An error severity assessed at levels 0 and 1 are considered to be ``recoverable,'' i.e., a communication partner would be able to understand what a speaker meant to say. An error level 2 represents a communication breakdown, where the perceived meaning is lost or drastically altered from a speaker's intent.

\begin{table}[t]
\begin{tabular}{|l|l|l|}
 \hline
   \textbf{ Type} & \textbf{Description} & \textbf{\# Errors (\%)}\\
    \hline
    Deletion & One or more spoken words & 413 (12\%) \\
    & do not appear in prediction. & \\
    \hline
    Contraction & Words either contracted & 17 (0.5\%) \\
    & or a contraction expanded & \\
    \hline
    Normalization & Non-canonical transcription & 404 (12\%) \\
    &(e.g. "four o'clock" vs "4:00")&  \\
    \hline
    Homophone & Word has same pronunciation  & 34 (1\%) \\
    & but different meaning. &\\
    \hline
    Spelling & Different spelling, beyond  & 30 (1\%) \\
    & what’s covered above. & \\
    & (e.g. “color” vs “colour”) &\\
    \hline
    Proper noun & Misrecognized named entity  & 386 (11\%) \\
    & or technical term.  & \\
    \hline
    Repetition & Non-spoken repetitions. & 21 (1\%) \\
    \hline
    Word Error & A word is misrecognized. & 2168 (62\%) \\
    &  (no above errors apply) & \\
    \hline
\end{tabular}
\caption{Description of error types with counts and proportion of 3473 errors.}
  \label{tab:error_types}
\end{table}

\begin{table}[h]
\centering
\begin{tabular}{|l|l|l|}
    \hline
    \textbf{Assessment} & \textbf{Description} & \textbf{\# Errors (\%)} \\
    \hline
    0 & Meaning is completely & 861 (25\%) \\
    & preserved. & \\
    \hline
    1 & Some errors, but meaning  & 786 (23\%) \\
    & is mostly preserved. & \\
    \hline
    2 & Major errors, significant  & 1826 (53\%) \\
    & changes to the meaning. & \\
    \hline
  \end{tabular}
  \caption{Error severity assessment response scale, descriptions, counts and proportion of total 3473 errors.}
  \label{tab:error_assessment scale}
\end{table}

\textbf{Dataset.} We consolidated the errors and assessments into a dataset. Table~\ref{tab:severity_counts} shows the distribution of the speakers and the severity of their speech impairment, Table~\ref{tab:error_types} shows distribution over error types and Table~\ref{tab:error_assessment scale} includes the assessment at different severity levels.
As shown in Table~\ref{tab:error_assessment scale}, 53\% the errors were assessed as Level 2 errors, 25\% of the transcription errors were marked as Level 0 (meaning completely preserved) and another 23\% as Level 1 (minor errors): in almost half of the cases, the meaning was preserved. The majority of errors were the most general category \textit{Word Error}. There were, however, a significant amount of normalization errors, which are typically meaning preserving (Levels 0 or 1). Examples of different error types and the associated SLP assessments are shown in Table~\ref{tab:error_examples}.

\begin{table*}[h]
\centering
\resizebox{\textwidth}{!}{
\begin{tabular}{lcrlll}
    \hline
    Error Type & Predicted Transcript & Actual Transcript & Word Acc. & F\textsubscript{BERT} & Assessment \\
    \hline
    Deletion & Come right back \_ & Come right back please & 0.75 & 0.86 & 0 \\
     & I have a \textit{head}\_ & I have a headache & 0.75 & 0.69 & 2 \\
    \hline
    Contraction &  \textit{I'm} a bit overwhelmed & I am a bit overwhelmed. & 0.60 & 0.89 & 0 \\
    \hline
    Normalization & play \textit{Beyoncé} & play Beyonce & 0.50 & 1.00 & 0 \\
    &  Okay \textit{9:30 five} & Okay, nine thirty five. & 0.50 & 0.75 & 1 \\
    \hline
    Proper Noun & Here are TV shows by Hugh \textit{Griffiths} & Here are TV shows by Hugh Griffith & 0.86 & 0.96 & 0 \\
    & \textit{First} do you know how the story ends & Faust, do you know how the story ends? & 0.88 & 0.79 & 2 \\
    \hline
    Repetition &  What \textit{are you} are you trying to say to me & What are you trying to say to me? & 0.75 & 0.92 & 1 \\
    \hline
  \end{tabular}
  }
  \caption{Examples of errors with associated Word Accuracy, F\textsubscript{BERT} and Error Assessment metrics.}
  \label{tab:error_examples}
\end{table*}

\subsubsection{Overview of transcript error assessment task}
\label{sub_sec:overview_of_task}
For completeness, we also detail the specific instructions provided to the SLPs here. %
SLPs were given the following instructions when assessing an error. To prevent bias and better simulate being a conversational partner, SLPs were asked to look at ground truth transcripts only after making a mental guess at the phrase meaning.

\begin{enumerate}
\item Read the model prediction without looking at the original transcript.
\item Make a mental guess at the intended meaning of the text.
\item Read the ground-truth transcript and compare to the prediction; assign an error assessment according to the scale in table.
\end{enumerate}

To assess Inter-Annotator Agreement, measured with Cohen's kappa~\cite{Cohen1960}, 5\% of the errors were annotated by two SLPs. Raters were in substantial agreement for both error assessment ($\kappa = 0.64$) and error types ($\kappa = 0.69$) based on the scale in \cite{Landis1977}.

\subsection{Word Accuracy and F\textsubscript{BERT}}

BERTScore~\cite{zhang2019bertscore} is a text evaluation metric based on a pretrained BERT model's~\cite{devlin2018bert} contextual embedding. It measures the similarity of two sentences as the sum of the cosine similarities of the token embeddings of the sentences as follows. Let the ground truth transcript ($x$) with pre-normalized token embeddings ($x_i$), be represented as $x = \langle x_1, \cdots, x_k \rangle $, and a similarly tokenized predicted sentence be represented as $y = \langle y_1, \cdots, y_l \rangle$, and the cosine similarity between two tokens $x_i$ and $y_j$ be denoted as $x_i \cdot y_j$. We use the version of BertScore where the cosine similarity is weighted by the inverse document frequency (idf) computed on the test set of ground truth transcripts. Then, BERTScore maximizes the cosine similarity of each token in $x$ with each token in $y$ to compute the recall, precision, and F1 measure as:
\begin{flalign}\label{eq:fbert}
    &R = \frac{1}{|x|}\sum_{x_i \in x} \max_{y_j \in y}(x_i \cdot y_j) 
    \quad
    P = \frac{1}{|y|}\sum_{y_j \in y} \max_{x_i \in x}(x_i \cdot y_j) \nonumber \\
       &F_{BERT} = 2 \frac{P \times R}{ P +  R} 
\end{flalign}

We use the BERT-base model~\cite{devlin2018bert} for the token embeddings to compute cosine similarity. BERTScore\cite{zhang2019bertscore} was evaluated on several hundred machine translation and captioning systems and was shown to be robust to syntax changes and provide a metric of semantic similarity that aligned well with human judgements. In this analysis we use the F1 measure, denoted F\textsubscript{BERT} (Eqn.~\ref{eq:fbert}), combining the sentence level precision and recall returned from BERTScore. 
The ranges of F\textsubscript{BERT} and WER\footnote{WER is capped at 100 in this analysis.} are [0,1] and [0,100] respectively. A transcript with an exact match would have F\textsubscript{BERT} = 1 and WER = 0. In order to make analysis more comparable, we match ranges by reporting Word Accuracy (Eqn. \ref{eq:word_acc_formula}) instead of WER in the subsequent sections. 
\begin{eqnarray}
\text{Word Accuracy} = 1 - \frac{WER}{100}
\label{eq:word_acc_formula}
\end{eqnarray}

\subsection{Ordinal Logistic Regression}

We examine how well WER and F\textsubscript{BERT} as metrics are aligned with expert-annotated error assessments.
Since error assessments were graded on a Likert Scale from 0 to 2 and are inherently ordinal, we ran a series of ordinal logistic regression (OLR) models with error assessments as the dependent variable, and Word Accuracy and/or F\textsubscript{BERT} as the independent variables.

To further determine model quality with Word Accuracy and F\textsubscript{BERT} as parameters, we rank each models' associated Akaike Information Criterion (AIC)\cite{akaike1998information}, a lower AIC indicating a better model for the data.

\begin{figure*}[h!]
    \centering
    {\includegraphics[width=0.35\textwidth]{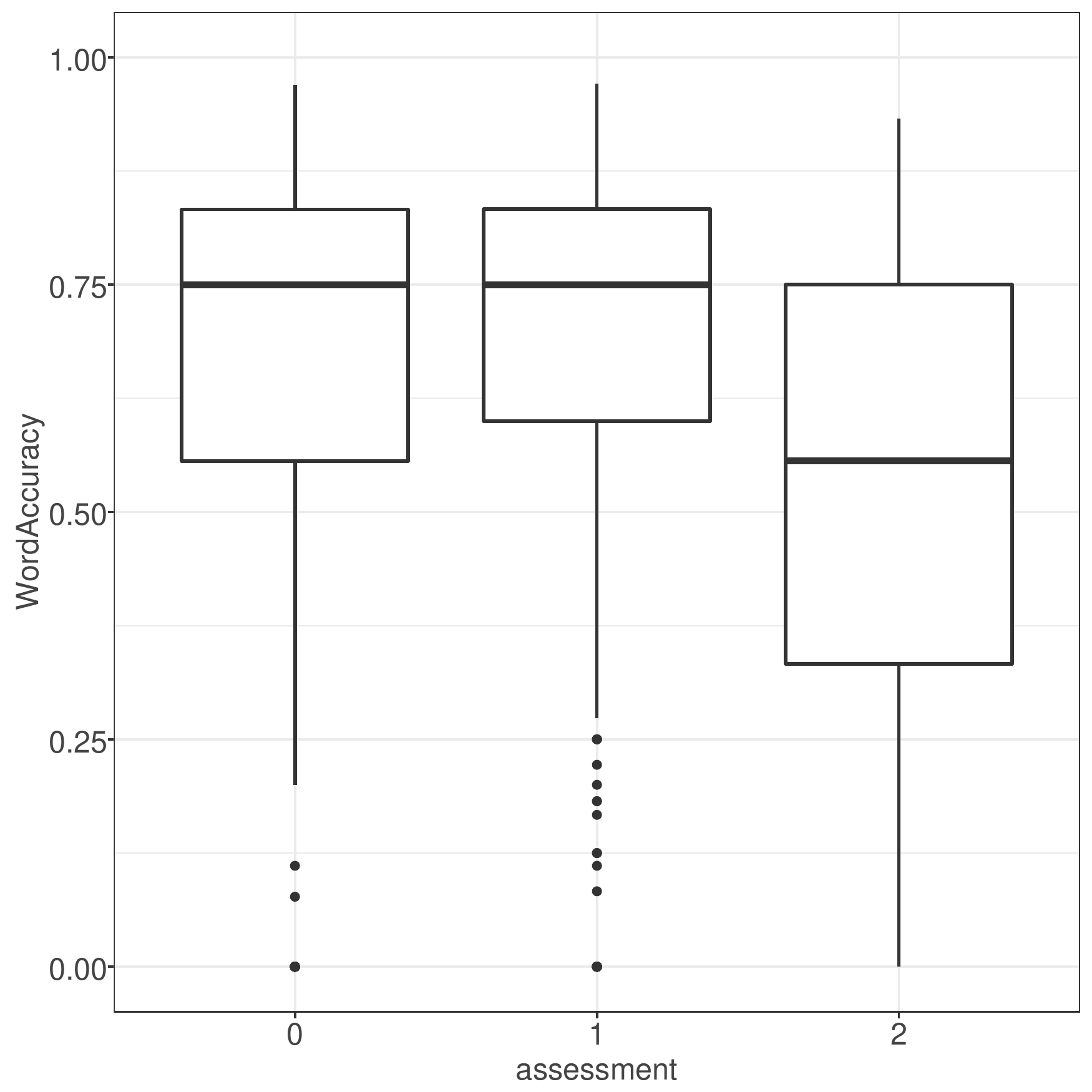}}\quad \quad
	{\includegraphics[width=0.35\textwidth]{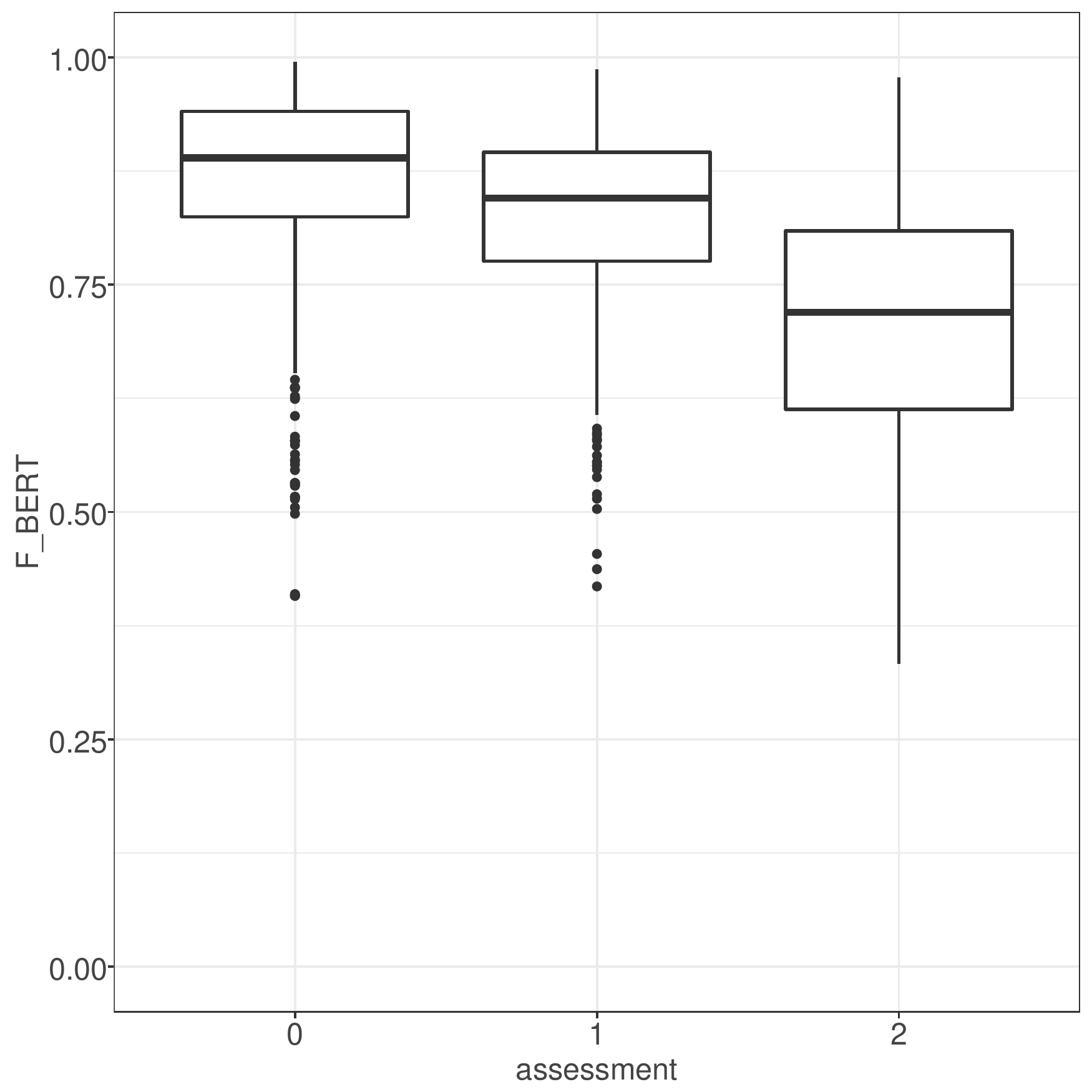}}\\
	\caption{Distribution of Word Accuracy (left) and F\textsubscript{BERT} (right) broken out by error assessment.}
	\label{fig:metric_v_asmt_boxplots}
\end{figure*}

\begin{figure*}[h!]
    \centering
    {\includegraphics[width=0.35\textwidth]{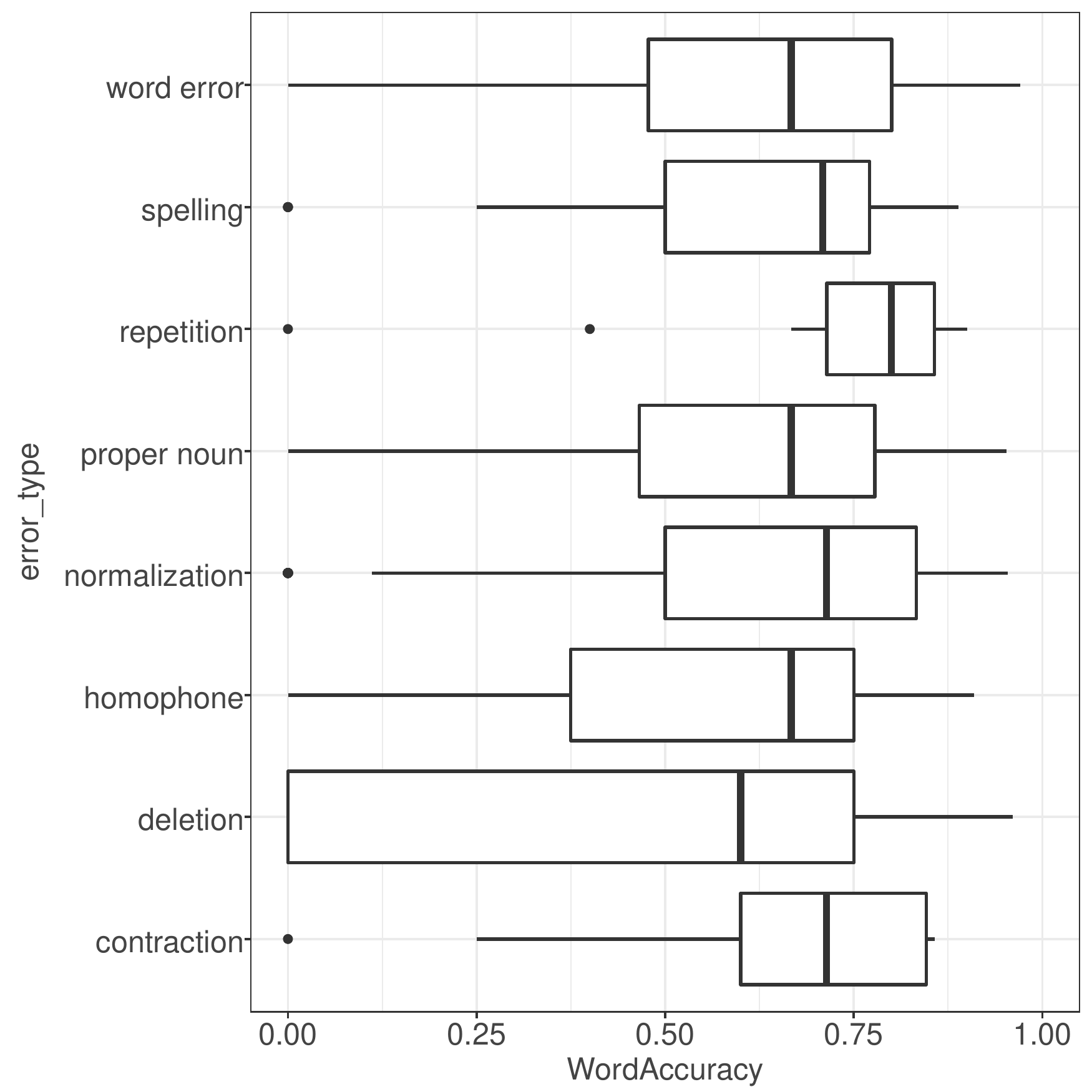}}\quad \quad
	{\includegraphics[width=0.35\textwidth]{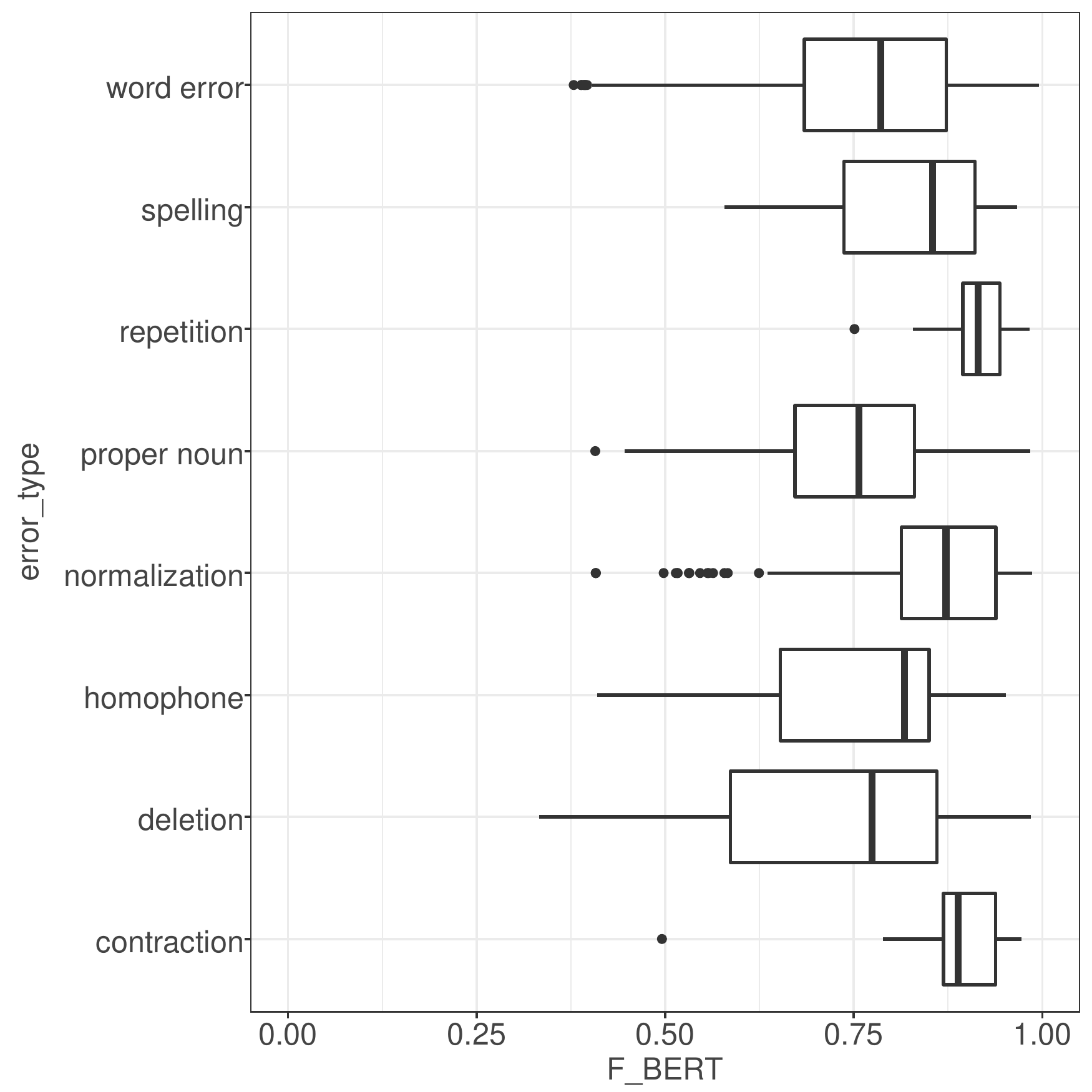}}\\
	\caption{Distribution of Word Accuracy (left) and F\textsubscript{BERT} (right) broken out by error type.}
	\label{fig:metric_v_type_boxplots}
\end{figure*}

\section{Results}
\label{sec:results}

\textbf{BERTScore distinguishes error severity better.} Figure \ref{fig:metric_v_asmt_boxplots} shows the summary statistics of Word Accuracy and F\textsubscript{BERT} for the three levels of error assessment. We can clearly see that 1) F\textsubscript{BERT} distinguishes different error assessment levels better than Word Accuracy, especially between 0 and 1; and 2) F\textsubscript{BERT} has overall smaller variance ($SD=0.142$) than Word Accuracy ($SD=0.274$). A one-way ANOVA test also confirms that F\textsubscript{BERT} ($F = 684.38, p<0.001$) could differentiate different assessment levels better than Word Accuracy ($F = 209.01, p<0.001$).

\textbf{BERTScore distinguishes error types better.} Leveraging the expert-annotated error types for all data, we then compare how Word Accuracy and F\textsubscript{BERT} differentiate between different error types, with results shown in Figure~\ref{fig:metric_v_type_boxplots}. For instance, F\textsubscript{BERT} differentiates contraction errors better than Word Accuracy (i.e. they should have higher scores because these errors typically do not impact the semantic meanings of the speech), similarly for normalization errors. One-way ANOVA tests again confirm that F\textsubscript{BERT} ($F = 41.75, p<0.001$) outperforms Word Accuracy ($F = 9.45, p<0.001$) when distinguishing different error types on impaired speech.

\textbf{BERTScore fits SLP assessments better.} Our results of ordinal logistic regressions show that both Word Accuracy and F\textsubscript{BERT} are significant predictors for error assessment, but F\textsubscript{BERT} has a larger absolute value of coefficient ($\text{coeff}=-10.87, t=-28.6, p<0.001$) than Word Accuracy ($\text{coeff}=-2.52, t=-17.4, p<0.001$), which means F\textsubscript{BERT} is more predictive of error assessment: a decrease in F\textsubscript{BERT} is more likely to result in a higher error assessment score (i.e. more severe error).
This is also confirmed by AIC scores where $AIC=5854$ for OLR with F\textsubscript{BERT} as the parameter and $AIC=6733$ with Word Accuracy. The OLR model with F\textsubscript{BERT} as the parameter fits better with the error assessment data, than with Word Accuracy as the parameter.

\textbf{BERTScore is more useful when measuring semantic similarity.}
Models based on WER and BERTScore both significantly fit the data, but there are situations where BERTScore is more robust to transcription errors that do not affect the semantic meanings. Table~\ref{tab:error_examples} shows transcription error examples with their associated Word Accuracy and F\textsubscript{BERT}. For instance, both ``Deletion'' examples (see row 1 in Table ~\ref{tab:error_examples}) have a word accuracy of 0.75. However, one (\textit{Come right back \_}) has an assessment score of 0 because the original meaning of the utterance is persevered, while the other example (\textit{I have a head\_}) has an assessment score of 2 because of the significant change to the intended meaning. Using Word Accuracy in this case fails to distinguish the quality of the predicted transcripts. F\textsubscript{BERT}, instead, performs better in measuring the quality of the predictions (0.86 vs. 0.69).
Similarly, normalization errors often result in low Word Accuracy, especially for short phrases, even though the meaning is not affected. E.g. \textit{play Beyoncé} vs. \textit{play Beyonce} has an Word Accuracy of 0.50 while an F\textsubscript{BERT} of 1.00. 

\textbf{BERTScore is however more expensive and less flexible than WER.}
WER is useful in many applications regardless of language or phrase domain. BERTScore on the other hand is more computationally expensive and relies on a pre-trained BERT model that can run into issues with multilingual data and out of vocabulary words.
\section{Conclusion}
\label{sec:conclusion}
ASR models exhibit higher word error rates when measured on non-typical speech, relative to rates typically reported in SOTA research. When making product decisions, it can be difficult to assess the quality of a model with a large WER. Models that could provide utility to people with impaired speech may be gated from release unnecessarily. To our knowledge, this is the largest user study validating the usefulness of personalized models despite models having WERs larger than products made for typical speakers. 

In this study we discussed a separate metric for determining model quality based on loss of semantic meaning in transcription errors using BERTScore, a semantic embedding metric. We found that BERTScore was more robust to several types of transcription errors including normalization and contraction errors. 

In future studies, we aim to investigate further metrics, such as Character Error Rate (CER) and BLEURT, that can give insight into when a model will be useful for speakers with disordered speech, especially when a speaker may have a condition that causes their voice to become more severely impaired over time. WER and BERTScore, when used together as parameters, fit the data better than either parameter alone. %
Future studies should also look at utilizing both signals together to assess model usefulness. %
It would also be good to a domain specific threshold of WER and F\textsubscript{BERT} that can help predict success rate, proportion of speakers that will have a usable ASR model after personalization, similar to the home automation domain WER of 15 discussed in Tobin and Tomanek(2022).

Future studies can also investigate applications of BERTScore for estimating ASR model quality beyond disordered speech. ASR models with higher WER may still be useful for speakers with accented speech or non-ideal acoustic settings for example.

\bibliographystyle{IEEEtran}
\bibliography{mybib}

\end{document}